\documentclass[apjl]{emulateapj}
\usepackage{amsfonts,amsmath,graphicx,natbib,apjfonts,subfigure}

\def\swift{{\it Swift}}

\def\cfa{1}
\def\atca{2}
\def\uva{3}
\def\col{4}
\def\israel{5}
\def\normale{6}
\def\inaf{7}
\shorttitle{GRB\,100316D}
\shortauthors{Margutti et al.}

\begin{document}
\title{The signature of the central engine in the weakest relativistic explosions: GRB\,100316D}
\author{R.~Margutti\altaffilmark{\cfa}, A.~M.~Soderberg \altaffilmark{\cfa}, 
M.~H. Wieringa\altaffilmark{\atca}, P.~G. Edwards\altaffilmark{\atca}, R.~A.~Chevalier\altaffilmark{\uva}, 
B.~J. Morsony\altaffilmark{\col}, R. Barniol Duran\altaffilmark{\israel}, L. Sironi\altaffilmark{\cfa}, B.~A. Zauderer\altaffilmark{\cfa}, D. Milisavljevic\altaffilmark{\cfa}, A. Kamble\altaffilmark{\cfa}, E. Pian\altaffilmark{\normale,\inaf}}

\altaffiltext{\cfa}{Harvard-Smithsonian Center for Astrophysics, 60 Garden St., Cambridge, MA 02138, USA}
\altaffiltext{\atca}{CSIRO Astronomy \& Space Science, Australia Telescope National Facility, P.O. Box 76, Epping, NSW 1710, Australia}
\altaffiltext{\uva}{University of Virginia, Astronomy Department, Charlottesville, VA 22904, USA}
\altaffiltext{\col}{Department of Astronomy, University of Wisconsin-Madison, 2535 Sterling Hall, 475 N. Charter Street, Madison WI 53706
-1582, USA}
\altaffiltext{\israel}{Racah Institute for Physics, Edmund J. Safra Campus, HebrewUniversity of Jerusalem, Jerusalem 91904, Israel}
\altaffiltext{\normale}{Scuola Normale Superiore, 7, I-56126 Pisa, Italy}
\altaffiltext{\inaf}{INAF - IASF Bologna, via Gobetti 101, I-40129 Bologna, Italy}

\begin{abstract}
We present late-time radio and X-ray observations of the nearby sub-energetic 
Gamma-Ray Burst (GRB)100316D associated with supernova (SN) 2010bh. 
Our broad-band analysis constrains the  explosion properties of 
GRB\,100316D to be intermediate between highly relativistic, collimated GRBs
and the spherical, ordinary hydrogen-stripped SNe. We find that $\sim 10^{49}$ erg
is coupled to mildly-relativistic ($\Gamma=1.5-2$), quasi-spherical ejecta, expanding 
into a medium previously shaped by the progenitor mass-loss with rate $\dot M
\sim 10^{-5}\,\rm{M_{\sun}yr^{-1}}$ (for wind velocity $v_{w}=1000\,\rm{km\,s^{-1}}$).
The kinetic energy profile of the ejecta argues for the presence of a central engine
and identifies GRB\,100316D as one of the weakest central-engine driven explosions
detected to date. Emission from the central engine is responsible for an excess of soft X-ray radiation 
which dominates over the standard afterglow at late times ($t>10$ days). 
We connect  this phenomenology with the birth of the most 
rapidly rotating magnetars. Alternatively, accretion onto a newly formed black hole 
might explain the excess of radiation. However, significant departure from the standard
fall-back scenario is required. 
\end{abstract}

\section{Introduction}
\label{Sec:Intro}

Gamma-Ray Bursts (GRBs, \citealt{Klebesadel73}) are the most powerful stellar explosions 
in our Universe, typically releasing $\sim10^{51}\,\rm{erg}$ (e.g. \citealt{Frail01}) 
coupled to highly relativistic jets. Long GRBs, with a duration of the
prompt $\gamma$-ray emission $\Delta t>2$ s \citep{Kouveliotou93},
are associated with the death of massive stars (see \citealt{Hjorth12}
for a recent review) and  give rise to the brightest displays. 
However, in the past few years, a new class of sub-energetic long GRBs has been recognized.
(\citealt{Soderberg06b} and references therein).  

Sub-energetic GRBs appear to be quasi-spherical explosions energetically
dominated by the non-relativistic ejecta that carry $\sim 99.9$\% of the 
explosion energy. Only $\sim0.1$\% of the energy is coupled to mildly
relativistic material. While the
mildly-relativistic ejecta clearly differentiate sub-energetic GRBs from ordinary  
hydrogen-stripped (Type Ib/c) SNe (\citealt{Soderberg10}), their relativistic energy release is $\sim 100-1000$
times lower than the typical $10^{51}$ erg of GRBs with fully 
relativistic outflows (\citealt{Frail01}), and a factor $\sim10^4$
lower than the most energetic GRBs \citep{Cenko11}. The energy coupled with the 
slow, non-relativistic ejecta is however interestingly similar between ordinary and sub-energetic 
GRBs and comparable to the most energetic Type Ib/c SNe (e.g. \citealt{Cano13}).
Sub-energetic GRBs therefore represent an intermediate class of explosions,
bridging the gap between the highly-relativistic, 
collimated GRBs and the more common Type Ib/c SNe.

Only a handful of sub-energetic GRBs have been discovered to date,
including the nearby GRBs 980425 (e.g. \citealt{Galama98,Kulkarni98}), 031203 
(\citealt{Soderberg04,Malesani04}) and 060218 (e.g. \citealt{Soderberg06b,
Campana06,Pian06,Mazzali06}). 
While the intrinsic faintness limited their detection to the local universe, 
the rate per unit volume of sub-energetic GRBs indicates that they are
$\sim10$ times more common than cosmological GRBs (\citealt{Soderberg06b,
Cobb06,Guetta07,Soderberg10}), which led some authors to conclude that
ordinary and sub-energetic GRBs have different origins  (e.g. \citealt{Virgili09}, \citealt{Bromberg11}). 
Recent simulations of jet-driven stellar explosions 
pointed instead to the possibility that the observed rates of events might reflect the
distribution of the duration of the central engine activity that powers the jet 
\citep{Lazzati12}. 
Ordinary GRBs would be produced by the longer-lasting engines able to launch
successful jets, while sub-energetic 
GRBs would result from \lq\lq failed\rq\rq  jets that barely pierce through the stellar surface,
thus connecting ordinary and sub-energetic GRBs to the same \lq\lq family\rq\rq of
explosions. The origin of sub-energetic GRBs and  their connection
to both ordinary Type Ib/c SNe and highly relativistic GRBs  is still an open issue.

More recently, the \emph{Swift} satellite (\citealt{Gehrels04}) added two new bursts
to the sub-energetic class:  GRBs 100316D (\citealt{Starling11,Fan11}) and
120422A (\citealt{Zhang12}, \citealt{Melandri12}, Zauderer et al., in prep).
The nearby ($z=0.0593$, \citealt{Chornock10}) 
GRB\,100316D triggered \emph{Swift} on 2010 March 16 at 12:44:50 UT,
which we use as explosion date throughout the paper. The smooth, long 
($\Delta t>1300\,\rm{s}$) and soft (spectral peak energy $E_{\rm{pk}}\sim20\,\rm{keV}$) 
prompt emission phase of GRB\,100316D is reminiscent of GRB\,060218 and it has been
studied by \cite{Starling11} and \cite{Fan11}. Spectroscopic follow-up
showed the emergence of clear SN features several days after trigger. The associated
SN, named SN\,2010bh, exhibited spectral features indicating a hydrogen-stripped progenitor and evolved 
similarly to previous GRB-SNe (\citealt{Chornock10,Cano11,Olivares12,Bufano12}).

Here we present the results from a broad-band, late-time monitoring of the sub-energetic
GRB\,100316D at X-ray and radio wavelengths.  Our observations 
constrain the explosion energy, 
its geometry and the properties of its environment, previously 
shaped by the progenitor mass-loss. We find evidence for a quasi-spherical,
central-engine driven explosion with mildly relativistic ejecta. Differently from ordinary GRBs, 
emission from the central engine of GRB\,100316D dominates over the standard
afterglow at late-times, allowing us to discuss its properties in the context
of black-hole and magnetar progenitors.

Throughout the paper we use the convention $F_{\nu}(\nu,t)\propto
\nu^{-\beta}\,t^{-\alpha}$, where the spectral energy index is related
to the spectral photon index by $\Gamma=1+\beta$.  Uncertainties
are quoted at 68\% confidence level, unless otherwise noted.
Standard cosmological parameters have been employed: $H_{0}=71$ km s$^{-1}$ Mpc$^{-1}$,
$\Omega_{\Lambda}=0.73$, and $\Omega_{\rm M}=0.27$.
\section{Observations}
\label{Sec:Obs}
\subsection{X-rays}
\label{SubSec:Xrays}

\begin{figure}
\vskip -0.0 true cm
\centering
\includegraphics[scale=1]{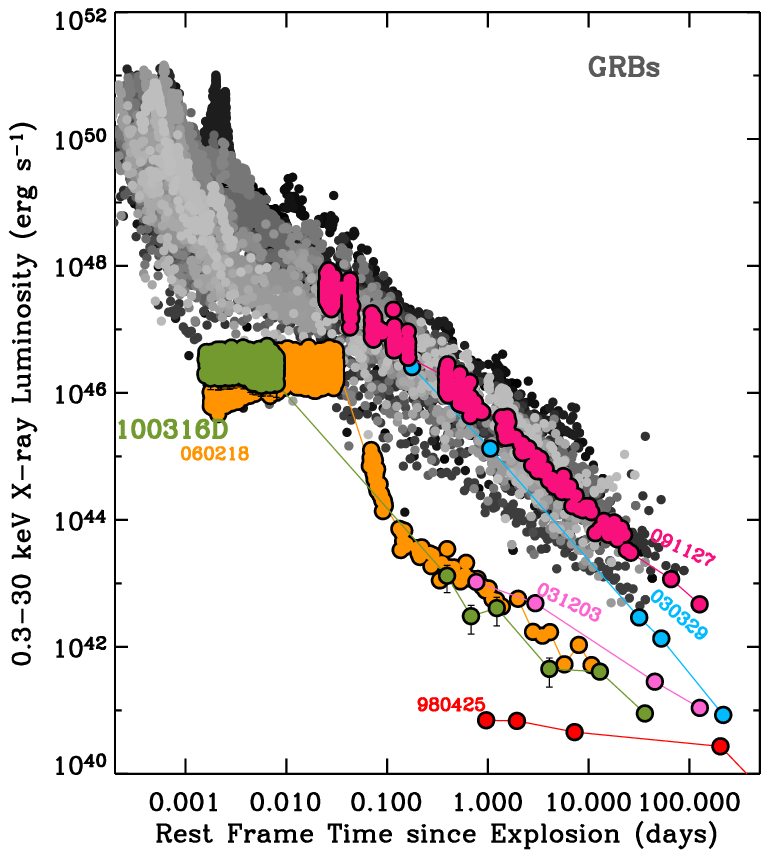}
\caption{X-ray luminosity of GRB\,100316D (including \emph{Swift}-XRT and \emph{Chandra}) compared to the
sample of 165 long GRBs with redshift observed by  \emph{Swift}-XRT  between 2004 and the
end of 2010, in the common rest-frame energy band 0.3-30 keV from \cite{Margutti13}. To this sample 
we add: GRB\,980425 (\citealt{Pian2000,Kouveliotou04}), GRB\, 031203 (\citealt{Watson04})
and GRB\,030329 (\citealt{Tiengo04}). GRBs with spectroscopically associated
SNe are in color and labeled. GRB\,100316D is similar to GRB\,060218 both at early and at late times.
Around $t\sim 40 $ days, the X-ray luminosity of GRB\,100316D approaches the level of GRB\,980425,
the least luminous X-ray afterglow ever detected.}
\label{Fig:Xray}
\end{figure}

\begin{figure}
\vskip -0.0 true cm
\centering
\includegraphics[scale=0.6]{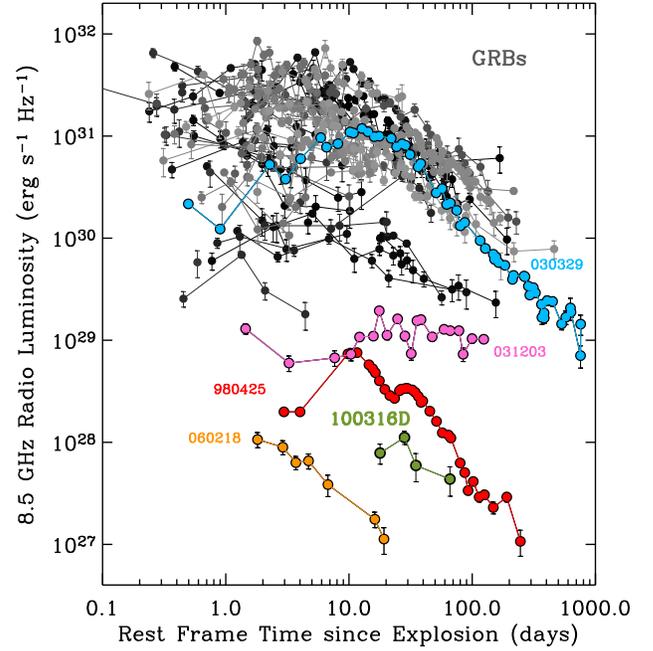}
\caption{8.5 GHz (rest-frame) radio afterglow of GRB\,100316D compared to the radio-selected sample 
of 129 GRBs observed between 1997 and  the end of 2011 for which a 
redshift measurement is available, presented in \cite{Chandra12}. GRBs with
spectroscopically associated SNe are in color and labeled. With a luminosity between 
GRB\,060218 and GRB\,980425 at $t\sim 10-70$ days, GRB\,100316D competes with 
the least luminous GRB radio afterglows ever detected.}
\label{Fig:radio}
\end{figure}

The \emph{Swift} X-Ray Telescope (XRT, \citealt{Burrows05}) started observing GRB\,100316D
127.5 s after the trigger, revealing a flat and smooth X-ray light-curve (Fig. \ref{Fig:Xray}), reminiscent of
the sub-luminous GRB\,060218 (\citealt{Campana06}). The early-time ($\Delta t<0.1$ days) 
X-ray emission of GRB\,100316D was extensively studied by \cite{Starling11}, who report the
presence of a thermal black-body component, later questioned by  \cite{Fan11}.  Our main focus
is on the late-time ($\Delta t>0.5$ days) X-rays.

We analyzed the XRT data using the latest HEAsoft release (v6.13), with standard filtering 
and screening criteria, and generated the 0.3-10 keV light-curve following the procedure 
outlined in \cite{Margutti13}. The use of the latest XRT calibration files (v13) leads to a severe
reduction of the statistical significance associated with the thermal black-body component,
with a chance probability of $10^{-3}$ (using v13) vs. $10^{-10}$ (using v11), according to
the F-test. In their re-analysis of the early X-ray data, \citealt{Starling12} found a similar trend
(see their Table 2).
Given the strong dependence of the statistical significance of the black-body component
on the instrumental calibration and the very limited impact\footnote{From a black-body plus 
power-law spectral fit, the best fitting intrinsic absorption is
$\rm{NH_{z}}=(0.81\pm0.07)\times 10^{22}\,\rm{cm^{-2}}$. The black-body
only contributes $\sim1.5\%$ to the total 0.3-10 keV fluence.} on the intrinsic neutral hydrogen 
column density estimate $\rm{NH_{z}}$, in the following we adopt 
$\rm{NH_{z}}=(0.68\pm0.02)\times 10^{22}\,\rm{cm^{-2}}$, as
obtained from a simple power-law spectral fit to the data between 127 s and 737 s, 
where no spectral evolution is apparent. The best-fitting photon index is 
$\Gamma_x=1.42\pm0.02$ 
and the Galactic column density is 
$\rm{NH}=7.1\times 10^{20}\,\rm{cm^{-2}}$ (\citealt{Kalberla05}).

At late times the X-ray emission from GRB\,100316D significantly softens. A spectrum
extracted in the time interval $0.4-2.1$ days (rest-frame) is well fit by an absorbed
power-law model with exceptionally soft  photon index $\Gamma_x=3.49\pm0.26$. 
GRBs typically show at this epoch $\Gamma_x\sim2$ (\citealt{Margutti13}).
This uncommon spectral behavior was previously observed in GRB\,060218 (\citealt{Fan06b}).

Motivated by the long-lived and peculiar afterglow of previous nearby GRB-SNe we initiated 
deep X-ray follow up of GRB\,100316D with the \emph{Chandra} X-ray Observatory.
Observations were obtained on 2010 Mar 30.3 and Apr 23.8 UT ($\Delta t\approx 13.8$ and
38.3 days) for 14 and 30 ksec, respectively (Program 11500488; PI Soderberg).
\emph{Chandra} ACIS-S data were reduced with the {\tt CIAO} software package
(v4.3), using the calibration database CALDB
v4.4.2, and applying standard ACIS data filtering. In both observations
we detected a source coincident with the \swift/XRT and radio positions (see Sec. 
\ref{SubSec:Radio}) with significance $\sim 6\sigma$ according to {\tt wavdetect}. Using a $1.5''$
aperture, the source count-rate in the 0.5-8 keV range is 
$(8.1\pm2.6)\times 10^{-4}\rm{ct\,s^{-1}}$ and $(4.9\pm1.4)\times 10^{-4}\rm{ct\,s^{-1}}$
for the first and second epoch, respectively. Assuming the spectral parameters from the XRT
analysis, the count-rates translate to an unabsorbed 0.3-10 keV flux of $(5.0\pm1.3)\times 10^{-14}\rm{erg\,
cm^{-2}s^{-1}}$ (first epoch) and $(2.7\pm0.7)\times 10^{-14}\rm{erg\,cm^{-2}s^{-1}}$
(second epoch).

Figure \ref{Fig:Xray} shows the complete X-ray data-set, with observations from 100 s until 40 days
since the explosion. Compared with a representative sample of GRBs in the common rest-frame
0.3-30 keV band from \cite{Margutti13}, GRB\,100316D is among the least luminous, both
during the prompt and during the late-time afterglow phase.
At $\sim40$ days, GRB\,980425 and GRB\,100316D are the least luminous GRB explosions
ever detected.

\subsection{Radio}
\label{SubSec:Radio} 

\begin{deluxetable}{lrccc}
\tablecaption{ATCA observations of GRB\,100316D/SN\,2010bh}
\tablewidth{0pt}
\tablehead{
\colhead{Date} & &\colhead{Time (Obs. Frame)} & \colhead{$F_{\nu,5.4}$} & \colhead{$F_{\nu,9.0}$}  \\
\colhead{(UT)} & &\colhead{(days)} & \colhead{($\mu$Jy)} & \colhead{($\mu$Jy)} \\
}
\startdata
2010 Mar &18.35 & 1.81 & $< 78$  & $<120$ \\
2010 Mar &27.10 & 10.57 & $< 81$  & $< 135$ \\
2010 Apr &4.46  & 18.93 & $90\pm 20$     & $< 63$ \\
2010 Apr &15.40 & 29.87 & $128\pm 19$    & $81\pm 26$ \\
2010 Apr &22.50 & 36.97 & $68\pm 21$     & $< 72$ \\
2010 May &25.40  & 69.87 & $50\pm 16$     & $< 60$ \\
2010 Sep &6.77  & 174.24 & $<42$     & $< 60$ \\
\enddata
\tablecomments{Errors are $1\sigma$ and upper limits are $3\sigma$.}
\label{tab:atca}
\end{deluxetable}

We observed GRB\,100316D with the Australia Telescope Compact Array
(ATCA) from 2010 Mar 18.35 UT to  Sep 6.77 UT ($\Delta t\approx 1.8-174$ days)
using the Compact Array Broadband Backend \citep{Wilson11}.
All observations were carried out at 5 and 9 GHz, with a bandwidth of 2 GHz
and are reported in Table~\ref{tab:atca}. We used PKSB1934-638 for
flux calibration, while phase referencing was performed using
calibrator PKSB0742-56 at the improved position reported by \cite{Petrov11}.  
We reduced the data using the MIRIAD package \citep{Sault95}.  

No radio source was detected in our first observation at $ t\approx 1.8$ days,
 enabling a deep limit of $F_{\nu}\lesssim 78$ and $120~\mu$Jy
($3\sigma$ rms) at 5.4 and 9.0 GHz, respectively
\citep{Wieringa10}.  A similarly deep observation at $ t\approx
10.6$ days also revealed no counterpart at either frequency.  
However, on Apr 4.5 UT ($ t\approx 19.0$ days), we clearly detected a 5.4
GHz source within the XRT error circle at $6.5\sigma$ significance.
Fitting a point source model,
we measured an integrated flux density of $F_{\nu}\approx 90\pm
20~\mu$Jy.  A contemporaneous observation at 9 GHz constrained $F_{\nu}\lesssim 63~\mu$Jy
($3\sigma$).  From our observations on Apr 15.4, the radio source is located at
position, $\alpha\rm (J2000)=07^{\rm h}10^{\rm m}30.47^{\rm s},
\delta\rm (J2000)=-56^{\rm o}15'20.03''$ with an uncertainty of $\pm
0."4$  in each coordinate.  The radio afterglow of GRB\,100316D was detected at 5.4 GHz until
$t\approx 70$ days. Our monitoring reveals a temporal peak 
around $ 30$ days at $\sim5$ GHz. 

When compared to other GRB afterglows in
Fig. \ref{Fig:radio}, GRB\,100316D competes with the least luminous
radio afterglows ever detected, with a luminosity between the sub-energetic  GRBs 060218 
and 980425.
The timescale of the temporal peak is however much later than for GRB\,060218.
\section{Properties of the late-time X-ray emission}
\label{Sec:Xrays}

\begin{figure}
\vskip -0.0 true cm
\centering
\includegraphics[scale=0.55]{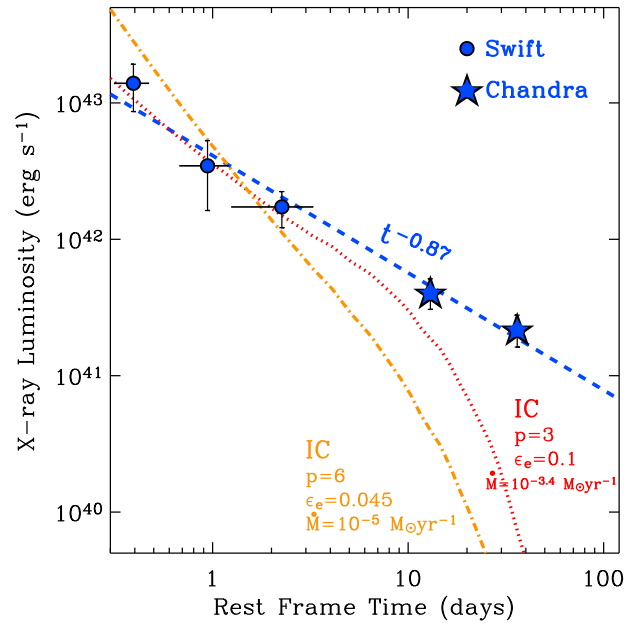}
\caption{Late-time evolution of the X-ray afterglow of GRB\,100316D as revealed
by \emph{Swift} and \emph{Chandra} observations in the 0.3-10 keV (observer frame)
energy band. The best-fitting power-law decay is $\propto t^{-0.87 \pm 0.08}$ (dashed blue line).
The expected contribution from Inverse Compton (IC) emission originating from up-scattered SN photospheric photons
by a population of electrons with $p=3$ (dot red line) and $p=6$ (dot-dashed orange line) that best fits the 
observations is also shown. For $t>10$ days, IC radiation substantially under predicts the observed 
luminosity and cannot explain the persistent late-time X-ray emission.}
\label{Fig:latetimeX}
\end{figure}

\begin{figure*}
\vskip -0.0 true cm
\centering
\includegraphics[scale=0.6]{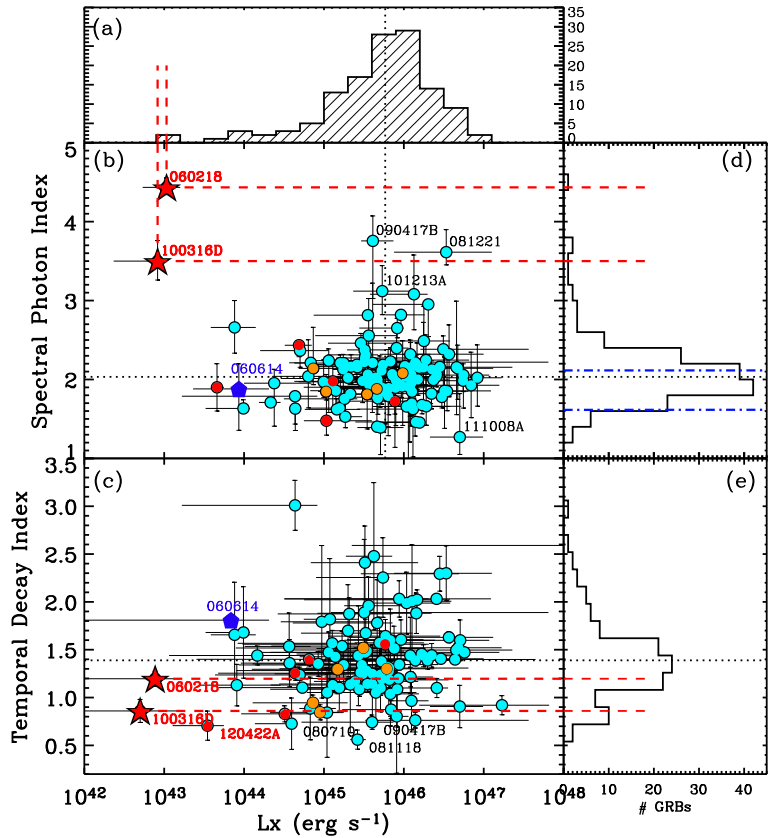}
\caption{Temporal and spectral properties of the late-time X-ray emission for a sample of 112 GRBs
detected by \emph{Swift} between November 2004 and July 2013, with known redshift. GRBs with spectroscopically
confirmed SNe are in red. GRBs with photometrically associated SNe are in orange. The blue diamond identifies
the SN-less GRB\,060614 (\citealt{DellaValle06,Fynbo06}). Dotted lines mark the median values of the distributions. \emph{Panel (b)}: 
X-ray spectral photon index $\Gamma_{x}$ computed in the rest-frame time interval 0.5-2 days as a function of the X-ray luminosity 
in that time interval (median value).  \emph{Panel (c)}: temporal decay index $\alpha_{x}$ as obtained from a local fit of the
X-ray light-curves between 0.5 and 10 days, rest-frame. \emph{Panels (a)-(d)-(e)}: projected distributions.
The dot-dashed lines in panel (d) mark the predicted $\Gamma_{x}$ that follows from the general
expectation of an electron power-law distribution $n_e(\gamma)\propto \gamma ^{-p}$ with $p\sim 2.23$
in relativistic shocks  with Fermi acceleration (\citealt{Kirk2000,Keshet05}), depending on whether the 
X-rays lie on the $F_{x}\propto \nu^{-p/2}$
(upper line) or the $F_{x}\propto \nu^{-(p-1)/2}$ (lower line) spectral segment.
GRBs 100316D and 060218 are clearly distinguished from all the other GRBs because
of their extreme spectral softness and low luminosity.  The decay of their X-ray light-curve is also
shallower than average.  }
\label{Fig:gammalpha}
\end{figure*}

\begin{figure}
\vskip -0.0 true cm
\centering
\includegraphics[scale=0.6]{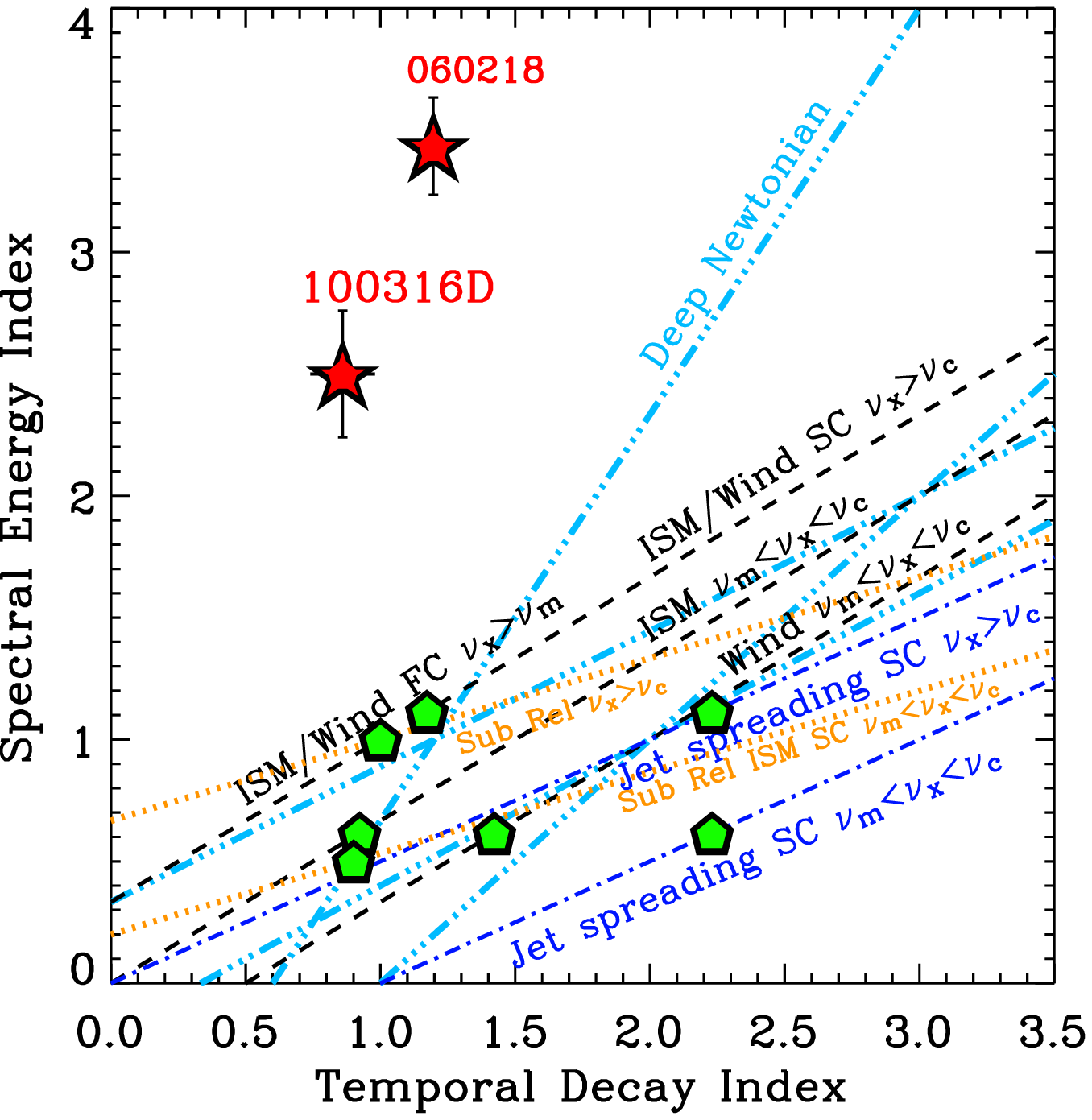}
\caption{Late-time (0.5-10 days) X-ray spectral energy index $\beta_x$ and temporal decay slope $\alpha_x$ 
of GRB\,100316D compared to the expectations from synchrotron radiation from a relativistic shock expanding 
into an ISM or wind-like medium (black dashed lines, \citealt{Zhang04,Zhang06}),
in the fast-cooling (FC) or slow-cooling (SC) regimes. $\nu_c$ refers to the synchrotron cooling frequency,
while $\nu_m$ is the characteristic synchrotron frequency.
Blue dot-dashed lines: 
uniform jet spreading with time (relativistic). 
We only plot closure relations for $p>2$ as indicated by our analysis in Sect. \ref{Sec:Xrays}.
Orange dotted lines: closure relations for a sub-relativistic shock (\citealt{Frail00}).
Light-blue, triple dot-dashed lines: expected $\beta_x-\alpha_x$ relations for a shock in the deep newtonian phase
(\citealt{Sironi13}).
Green diamonds mark the position of a blast wave with a power-law population of radiating electrons with 
$p=2.23$ (as expected for relativistic shocks, \citealt{Kirk2000,Keshet05}) or $p=2$ 
(general expectation for non-relativistic shocks, e.g. \citealt{Blandford87}).
GRBs 100316D and 060218 are  clearly \emph{not} consistent with these predictions.}
\label{Fig:closure}
\end{figure}

The late-time X-ray emission of GRB\,100316D is characterized by a mild decay $\propto t^{-\alpha_x}$
with index $\alpha_x=0.87 \pm 0.08$ (for $t>0.3$ days, rest frame, see Fig. \ref{Fig:latetimeX}), compared
to the steeper decay $\alpha_x\sim1.4$ typically observed in GRB afterglows at this epoch
(Fig. \ref{Fig:gammalpha}). The X-ray spectrum is unusually soft, with a power-law spectral energy index
$\beta_x=2.49\pm 0.26$. The low luminosity and exceptional spectral softness clearly distinguish
GRBs 100316D and 060218 from all the other GRBs and GRB-SNe of Fig. \ref{Fig:gammalpha}.
Independently from the X-rays being above or below the synchrotron cooling frequency $\nu_c$,
$\beta_x\sim2.5$ implies a very steep distribution of shocked electrons $n_e(\gamma)\propto \gamma^{-p}$,
with $p=5-6$. In the standard external forward shock model, such a steep electron distribution 
would naturally produce a fast-decaying light-curve with $\alpha_x>3.3$, significantly
steeper than the observed $\alpha_x\sim 0.9$ (Fig. \ref{Fig:closure}). 
The exceptionally soft spectrum and the milder than average temporal decay are therefore 
not consistent with the standard external forward shock origin.

The dynamics of the forward shock above would be however modified by continuous energy
injection by the central engine into the shock. Following \cite{Zhang06}, we consider  an 
injection luminosity term scaling as $L(t)=L_0(t/t_0)^{-q}$. The energy in the fireball
scales as $E\propto t^{1-q}$. 
The mild temporal decay of GRB\,100316D and its super-soft X-ray spectrum
would require an injection luminosity index $q<-0.3$,
implying a \emph{rising}
injected luminosity with time (\citealt{Zhang06}, their Table 2).  GRB progenitor
models either lead to a black hole torus system (e.g. 
\citealt{Narayan92,Woosley93,Paczynski98,Meszaros99,Fryer99}) 
or to a highly-magnetized, rapidly rotating pulsar (i.e. a magnetar, e.g. \citealt{Usov92,
Duncan92}). The initial spin-down luminosity from a millisecond pulsar 
requires $q=0$, evolving to $q=2$ at later times, when the electromagnetic dipolar radiation
dominates (e.g. \citealt{Dai98,Zhang01}). The injection luminosity term of a black hole torus system
is typically characterized by $q=5/3$ (or $q=4/3$) at late times (\citealt{MacFadyen01,Janiuk04}
and references therein), and hence has no impact on the dynamics of an adiabatic fireball 
(for which $q<1$ is required \citealt{Zhang01}).  A  study of X-ray flares in GRBs 
points to an even steeper index $q\sim2.7$ (\citealt{Margutti11b}) at earlier times.
For both a magnetar and a black hole system, the injected luminosity index is therefore $q\geqslant 0$,
not consistent with $q<-0.3$ required to explain the properties of the late-time X-ray emission
of GRB\,100316D. We conclude that the late-time X-rays are unlikely to originate from 
the emission from the explosion shock decelerating into the environment. This
conclusion is independently supported by the analysis of the contemporaneous
radio emission in Sec. \ref{Sec:RadioXrays}. Alternative scenarios 
are explored in Sec. \ref{Sec:nature}.

\section{Late-time radio-to-X-rays energy distribution}
\label{Sec:RadioXrays}

\begin{figure}
\vskip -0.0 true cm
\centering
\includegraphics[scale=0.55]{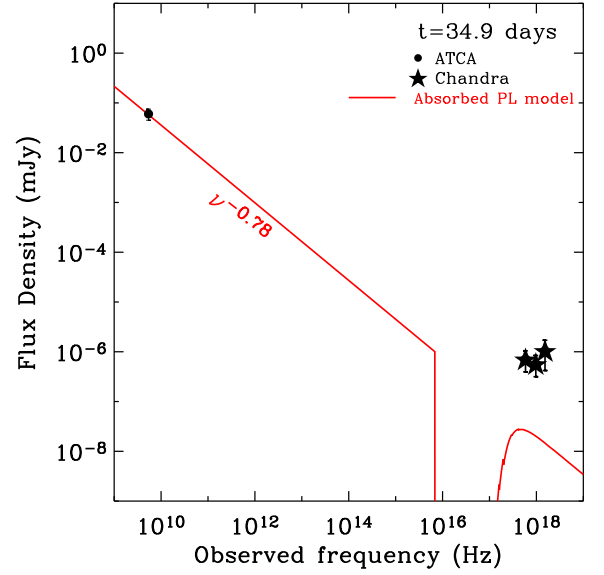}
\caption{Radio-to-X-ray SED at $t=34.9$ days (rest frame) revealing the
presence of an excess of X-ray radiation with respect to the expected 
synchrotron emission (red solid line). We show here
the most optimistic spectral model with $F_{\nu}\propto \nu^{-0.78}$. 
Radio observations at $t=17.9$ days (rest frame) indicate $\beta_R>0.78$. 
A spectral break between radio and X-rays would lead to an even
more pronounced X-ray excess, leading us to conclude that
the synchrotron forward shock model is unable to explain the bright late-time X-ray emission.}
\label{Fig:sed5}
\end{figure}

We study the radio-to-X-rays spectral energy distribution (SED) at 6 different epochs,
corresponding to the times of the radio observations in Table \ref{tab:atca}. 
At these times the optical wavelengths are dominated by the SN photospheric emission,
and are therefore not included here.
In the standard afterglow scenario, radio and X-ray photons  originate from 
the same population of electrons accelerated to relativistic speed by the explosion 
shock, and are expected to lie on the same synchrotron spectrum (e.g.
\citealt{Granot02}, their Fig. 1).  We test this prediction below.

An SED extracted at $t=1.7$ days (rest frame) constrains the radio to X-ray spectral 
index to an unusually flat value, $\beta_{RX}<0.6$. A similar value was observed
only in the case of GRB\,060218, for which $\beta_{RX}=0.5$ a few 
days after the explosion (\citealt{Soderberg06b}). At this time, the X-rays 
show an uncommon, exceptionally steep spectrum, with $\beta_X\sim 2.5$ (Sect. \ref{Sec:Xrays}
and Fig. \ref{Fig:gammalpha}). 

Further radio observations at $t=17.9$ days indicate a decreasing radio
flux density with frequency, $F_{R}\propto \nu^{-\beta_R}$ with $\beta_R>0.78$,
implying that, by this time, the spectral peak frequency has crossed the radio band
and $\nu_{\rm{sa}}\lesssim9$ GHz, (where $\nu_{\rm{sa}}$ is the synchrotron self-absorption
frequency). We use the constraint $\beta_R>0.78$ to compute an upper limit to the 
contemporaneous X-ray radiation originating from synchrotron emission. 
Figure \ref{Fig:sed5} shows the results at $t=34.9$ days (rest-frame).
The observed X-ray flux is significantly brighter than the extrapolation of the 
synchrotron model even in the most optimistic case of a simple power-law spectral 
distribution with $\beta=0.78$. We conclude that even if we ignore the exceptionally
soft X-ray spectrum,  the synchrotron model is unable to
explain the late-time X-ray emission in GRB\,100316D, which
should be interpreted as a different component (Sect. \ref{Sec:nature}). 
For $t>1$ day, X-ray and radio photons cannot be connected with the emission from a single
shock wave.  It is remarkable to note that the observed late-time X-ray and radio emission of 
GRB\,060218 shares these same properties and points to the same conclusion (\citealt{Soderberg06b,Fan06b},
see also \citealt{Waxman07}).

\section{Radio calorimetry and jet opening angle}
\label{Sec:RadioCal}

\begin{figure*}
\vskip -0.0 true cm
\centering
\includegraphics[scale=0.75]{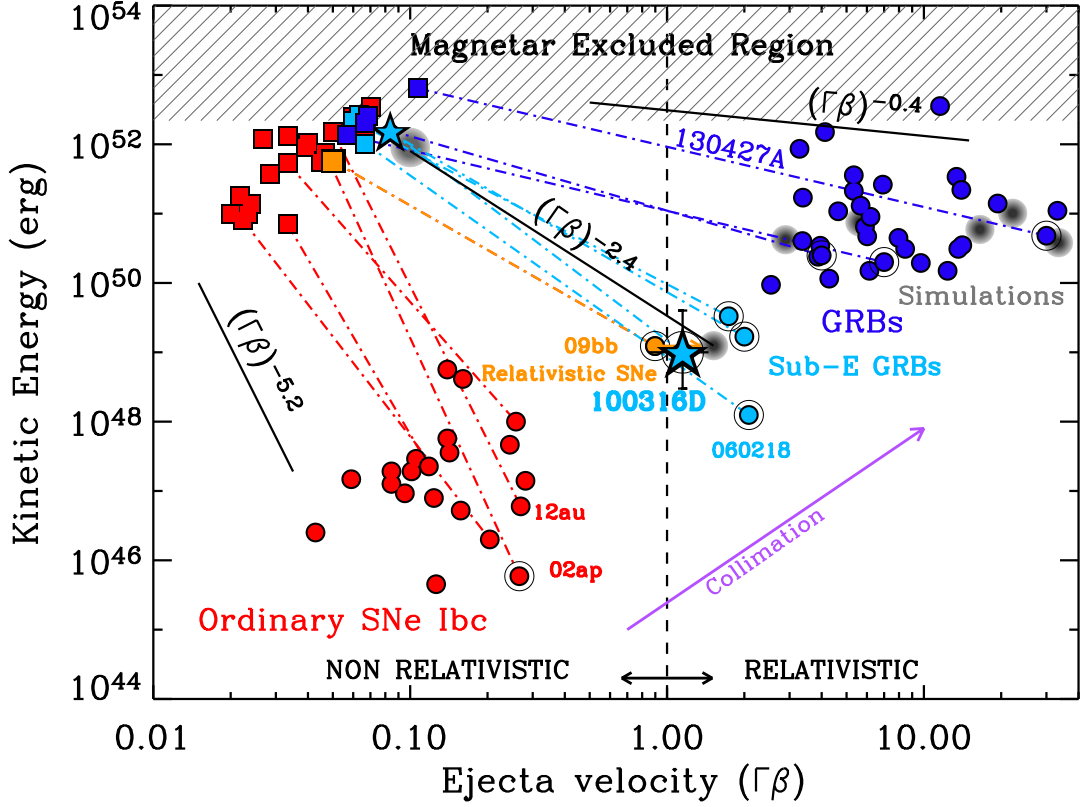}
\caption{Kinetic energy-velocity profile of the ejecta of ordinary SNe Ib/c (red), relativistic SNe (orange), 
sub-energetic GRBs (light blue) and GRBs (blue).  Squares: $E_k$ in slow moving material estimated
from the modeling of the SN optical emission. Circles mark the $E_k$ in the fastest ejecta
as measured from radio observations of SNe and broad-band afterglow modeling of GRBs, for which 
we report the beaming corrected values. For GRBs, $\Gamma\beta$ of the fast component is 
estimated at 1 day  rest-frame and includes deceleration of the blast-wave into the environment.
Open black circles identify explosions with broad lines 
in their optical spectra. Grey points: results obtained from simulations
of jet-driven explosions with energy $E=2\times10^{52}$ erg and different duration of the central engine
\citep{Lazzati12}.
Dashed-dotted lines connect measurements of the same explosion.  
Ordinary SNe are characterized by very steep profiles where a negligible fraction 
of energy is coupled to the fastest moving material, in general agreement with the expectations
from pure hydrodynamic collapse $E_k \propto (\Gamma \beta)^{-5.2}$, where no central engine is
involved  \citep{Tan01}. GRBs distribute their
energy budget differently, with comparable energy in their relativistic (or mildly relativistic)
and slow ejecta. The result is a much flatter $E_k$ profile, typical of jet-driven explosions 
with long-lasting central engines ($E_k\propto (\Gamma \beta)^{-0.4}$ for an explosion
energy of $2\times10^{52}$ erg and engine duration of  7 s,  \citealt{Lazzati12}).
Relativistic SNe and sub-energetic GRBs are intermediate ($E_k\propto (\Gamma \beta)^{-2.4}$), and fall into
the parameter space occupied by weak jet-driven explosions where the jet barely pierces through the 
stellar surface .\footnote{References:  Ordinary Type Ib/c SNe: \cite{Berger03a}, \cite{Soderberg06b},
\cite{Soderberg08}, \cite{Soderberg10}, \cite{Soderberg10b}, \cite{Sanders12}, \cite{Cano13}, 
\cite{Milisavljevic13}, \cite{Mazzali13}, Kamble et al., in prep.  Sub-energetic GRBs:  \cite{Berger03b}, \cite{Soderberg06c}
and references therein, \cite{Cano13}. Relativistic SN\,2009bb: \cite{Soderberg10}, \cite{Cano13}.
GRBs: \cite{Berger03a}, \cite{Frail06}, \cite{Chandra08}, \cite{Cenko10}, \cite{Cenko11}, 
\cite{Troja12}, \cite{Cano13}, \cite{Xu13}, \cite{Laskar13}, \cite{Perley13}, Guidorzi et al. in prep.,  Laskar et al. in prep.}}
\label{Fig:Ekvelocity}
\end{figure*}

Using the radio afterglow of GRB\,100316D we constrain  its kinetic energy, jet opening angle
and density of the environment.
Observations at $t_1\sim18$ days (rest-frame) indicate that the spectral peak frequency is just below
the radio band, $\nu_{\rm{sa}}(t_1)\lesssim9$ GHz, while the peak of the light-curve
around $ t_2\sim 30$ days suggests that $\nu_{\rm{sa}}(t_2)\sim5$ GHz, with
$F_{\nu_{sa}}\sim130\, \mu \rm{Jy}$. Applying the standard formulation of GRB
afterglows powered by synchrotron emission (\citealt{Granot02}, their Table 2)
and assuming a wind-like $n\propto r^{-2}$ environment  as appropriate for massive stars at the end of
their evolution, we constrain the fireball kinetic energy,  $E_{k}$, and the
progenitor mass-loss rate, $A_{\star}$ (defined following \citealt{Chevalier00}). 
For an electron power-law index $p=2.1-2.3$
and microphysical parameters  $\epsilon_e=0.01-0.1$, $\epsilon_B=0.01$, 
the fireball kinetic energy is $E_{k}=(0.3-4)\times 10^{49}\,\rm{erg}$ coupled to 
mildly relativistic ejecta with $\Gamma\sim 1.5-2$ (at 1 day rest-frame).
The mass-loss rate is 
$A_{\star}=0.4-1$ corresponding to  $\dot M=(0.4-1)\times10^{-5}\,\rm{M_{\sun}}
\,\rm{yr^{-1}}$ for wind velocity $v_w=1000\,\rm{km\,s^{-1}}$. Compared
with its cosmic twin GRB\,060218, GRB\,100316D is $\sim10$ times more energetic 
but exploded in a much denser environment\footnote{This suggests
that  the high $\rm{NH_{z}}\sim7\times10^{21}\,\rm{cm^{-2}}$ of  Sec.\ref{SubSec:Xrays}
is local to the explosion, as opposed to arising from material
that happens to be along our line of sight. For comparison, this value is a
factor $\sim2$ higher than the $\rm{NH_{z}}$ inferred for GRB\,060218 \citep{Soderberg06b}.}. The higher environment
density of GRB\,100316D explains why the blast-wave associated with the less energetic 
GRB\,060218 propagates with comparable but higher velocity:
\cite{Soderberg06b} infer $\Gamma\sim2.3$ at $t\sim 5$ days. 
The progenitor of GRB\,100316D suffered from more
consistent mass-loss before exploding,
causing the fireball to decelerate on shorter time scales compared
to shocks propagating in less dense environments like GRB\,060218,
for which $\dot M\sim \times10^{-7}\,\rm{M_{\sun}}\,\rm{yr^{-1}}$ (\citealt{Soderberg06b}).

The late-time temporal decay of the radio light-curve  $\propto t^{-\alpha_r}$
with $\alpha_r<1$ is much shallower than 
the $t^{-2.2}$ behavior expected after a jet break \citep{Sari99} and suggests
that the fastest ejecta responsible for the radio emission  is not strongly collimated.  
Our radio monitoring constrains the jet break time $t_j>66$ days (rest-frame)
which formally translates into a jet opening angle $\theta_j>80^\circ$, for the
kinetic energy and environment density determined above (see \citealt{Chevalier00}, their equation 31).
As for GRB\,060218 (for which $\theta_j>80^\circ$, \citealt{Soderberg06b}), 
the radio observations argue for a mildly relativistic explosion
with quasi-spherical ejecta. 

These properties place the sub-energetic GRB\,100316D between the
highly relativistic, collimated GRB explosions and the spherical, ordinary Type Ib/c
SNe as we show in Fig. \ref{Fig:Ekvelocity}. 
Its kinetic energy profile $E_k\propto (\Gamma \beta)^{-2.6}$ is significantly flatter\footnote{The
dense environment around GRB\,100316D significantly decelerated the
fastest ejecta by 1 day after the explosion, time at which we compute $\Gamma \beta$
in Fig. \ref{Fig:Ekvelocity}. The ``intrinsic''  $E_k (\Gamma \beta)$ profile of the explosion is therefore
flatter than what shown.} than that expected 
from a pure hydrodynamic collapse (where $E_k \propto (\Gamma \beta)^{-5.2}$, \citealt{Tan01}), 
and argues for the presence of a central engine able
to accelerate a non-negligible fraction of the ejecta to mildly relativistic speeds. A flatter
$E_k(\Gamma \beta)$ profile is expected for jet-driven stellar explosions \citep{Lazzati12},
with the least steep $E_k(\Gamma \beta)$ profiles associated with powerful jets able
to break out through the stellar surface while the engine is still active. Highly relativistic 
GRBs belong to this category of explosions and show
$E_k\propto (\Gamma \beta)^{-\delta}$ with $\delta <1$ (Fig. \ref{Fig:Ekvelocity}). GRB\,100316D
seems instead to be associated with the class of jet-driven explosions where the jet head is 
only barely able to reach the surface (or even breaks out after the end
of the central engine activity), resulting in an $E_k(\Gamma \beta)$ 
profile intermediate between a pure hydrodynamic collapse (where no jet is formed
at any stage of the collapse) and that of fully-developed, jet-powered explosions. 
\section{Nature of the late-time X-ray emission}
\label{Sec:nature}

GRB\,100316D is characterized by an unusually flat radio to X-ray spectral index,
an exceptionally soft late-time X-ray emission and a flatter than average
X-ray temporal decay (Sec. \ref{Sec:Xrays}). These properties led us to 
identify the presence of an X-ray excess of emission with respect to the
standard afterglow model powered by synchrotron radiation (Sec. \ref{Sec:RadioXrays}).
We discuss here the physical origin of the excess of emission and
conclude that this excess is connected to the explosion central engine.
\subsection{Inverse Compton emission}
\label{SubSec:IC}
Inverse Compton (IC) emission originating from the up-scattering of optical photons from 
the SN photosphere by a population of electrons accelerated to relativistic speeds by the shock wave
potentially contributes to the observed X-ray emission. The IC X-ray emission is negligible for
cosmological GRBs, but might be relevant for nearby bursts. In ordinary hydrogen-stripped 
SNe exploding in low-density environments, IC is the main X-ray emission mechanism
during the first $\sim40$ days after the explosion (\citealt{Bjornsson04,Chevalier06}).
We estimate the expected IC contribution to the X-ray afterglow of GRB\,100316D
by employing the formalism by \cite{Margutti12} modified to account for a SN ejecta outer density 
structure scaling as $\rho_{\rm{SN}}\propto R^{-n}$ with $n\sim10$, as appropriate
for Type Ib/c SNe (\citealt{Matzner99}). The IC X-ray luminosity depends on
the structure and density of the environment swept up by the blast wave $\rho_{\rm{CSM}}(R)$;
the details of the electron distribution $n_{e}(\gamma)=n_0\gamma^{-p}$ responsible for the up-scattering;
the fraction of shock energy in relativistic electrons $\epsilon_{e}$;
the explosion parameters (ejecta mass $M_{\rm{ej}}$ and kinetic energy $E_{\rm{k}}$) and
the SN bolometric luminosity ($L_{\rm{IC}}\propto L_{\rm{bol}}$).

The bolometric luminosity of SN\,2010bh has been computed by \cite{Bufano12}. Their modeling of 
the bolometric light-curve points to an energetic SN explosion with $E_{\rm{k}}\sim10^{52}\,\rm{erg}$
and $M_{\rm{ej}}\sim3\,\rm{M_{\sun}}$ (see also \citealt{Cano11}, \citealt{Olivares12} and \citealt{Cano13}
who obtained consistent results). Using these values and a wind-like 
CSM  ($\rho_{\rm{CSM}}\propto R^{-2}$, as expected from a star which has been losing material at constant rate $\dot M$),
we  show in Fig. \ref{Fig:latetimeX} the IC models that best fit our observations. We show
the results for both a population of radiating electrons with $p=3$, as typically indicated by radio observations of 
type Ib/c SNe (e.g.  \citealt{Chevalier06}), and for electrons with a much steeper
distribution with $p=6$, as suggested by the exceptionally soft late-time X-ray spectrum of GRB\,100316D.
The mass-loss rate, $\dot M$, is reported in Fig. \ref{Fig:latetimeX} for a wind velocity  $v_{w}=1000\,\rm{km\,s^{-1}}$,
typical of Wolf-Rayet stars.

Irrespective of the assumed electron density distribution, the IC X-ray luminosity declines steeply
after SN\,2010bh reaches maximum light around $t\sim10$ days (see \citealt{Olivares12}, their
Fig. 7), due to the substantial decrease of optical seed photons from the SN
photosphere. As a consequence, the IC mechanism severely under-predicts the observed
luminosity at late times (Fig. \ref{Fig:latetimeX}), failing to explain the persistent late-time 
X-ray emission in GRB\,100316D. The fully relativistic treatment of the IC emission 
by  \cite{Waxman07} leads to the same conclusion.
\subsection{Shock break out}
\label{SubSec:BO}

GRB\,060218 is the only other known explosion that shares with GRB\,100316D 
the same unusual properties of the late time X-ray and radio emission (Fig.
\ref{Fig:gammalpha}), including
the evidence for an excess of soft X-ray radiation (\citealt{Soderberg06b,Fan06b}).
Both explosions are also characterized by a peculiar $\gamma$-ray 
prompt emission phase consisting of a smooth, long ($\Delta t> 1300$ s) and
soft (time averaged spectral peak energy $E_{\rm{pk}}<20$ keV) pulse of emission
releasing a modest amount of energy $E_{\rm{iso}}\sim 5\times 10^{49}$ erg 
(\citealt{Kaneko07, Starling11}), setting these two bursts apart from all the
other GRBs.
The properties of these two bursts also sets them apart from the handful of known sub-energetic GRBs as well.
 It is thus reasonable to think that the physical origin of the unusual early-time and late-time properties is in some 
way related.

The prompt emission of GRBs 060218 and 100316D have been 
explained by \cite{Nakar12}\footnote{See also \cite{Waxman07} and \cite{Wang07}
for GRB\,060218.} as radiation from a relativistic shock breaking out at a large radius
$R_{\rm{bo}}\sim5\times 10^{13}$ cm (for GRB\,060218) and $R_{\rm{bo}}>6\times 10^{13}$ 
cm  (for GRB\,100316D, for which only a lower limit can be placed on the $\gamma$-ray
energy $E_{\rm{iso}}>6\times10^{49}$ erg due to an orbital data gap, \citealt{Starling11}). 
By $\sim 40$ days after the explosion, the typical temperature of the shock break out 
radiation $T_{\rm{bo}}$ is however significantly below the X-ray band
($T_{\rm{bo}}\ll0.1$ keV using the formalism by \citealt{Nakar10,Nakar12}), 
leading us to conclude that the contribution of residual radiation 
from the shock break out to the late-time X-rays is negligible.
\subsection{Long-lived Central Engine}
\label{SubSec:CE}

The $E_k(\Gamma \beta)$ profile of Sec \ref{Sec:RadioCal} (Fig. \ref{Fig:Ekvelocity}) 
argues for the presence of a central engine and identifies GRB\,100316D as one of the 
weakest central-engine driven explosions detected so far. 
We consider here the possibility of radiation originating from the explosion
remnant, in the form of a long-lived central engine. The same possibility was
considered by \cite{Fan06b} for GRB\,060218. 
The collapse of massive stars typically leads to a black hole plus long-lived 
debris torus system (\citealt{Narayan92,Woosley93,Paczynski98,Meszaros99,Fryer99}) 
or to a fast-rotating, highly-magnetized pulsar (i.e., a magnetar, e.g., \citealt{Usov92,
Duncan92}).  

Accretion onto a  black hole is a well known source of high-energy radiation.
From the observed $L_{x}\sim3\times 10^{41}\,\rm{erg\,s^{-1}}$ at $\sim30$ days after the explosion,
we estimate an accretion rate of $\dot M_{\rm{acc}}\sim10^{-10}-10^{-8}\,\rm{M_{\sun}s^{-1}}$
assuming an accretion efficiency $\eta_{\rm{acc}}=0.001-0.01$ (e.g. \citealt{MacFadyen01})
and an X-ray to bolometric flux correction $\eta_x=0.01-0.1$ (\citealt{Fan06b}).
The inferred range of $\dot M_{\rm{acc}}$ is consistent with the extrapolation
of the expected accretion rates from fall back calculated by \cite{MacFadyen01}
in the context of the collapsar model. This result would point to similar
late-time accretion rates between bright bursts (for which the model by \citealt{MacFadyen01} 
was originally developed) and sub-energetic GRBs. The observed temporal decay 
$L_{x}\propto t^{-0.87}$ is however significantly shallower than what expected in the context of
fall back accretion models that predict $L\propto t^{-5/3}$ (e.g. \citealt{Chevalier89}). While the mild $L_{x}$
decay does not rule out accretion as source of the late-time X-ray excess of radiation,
it does require substantial departure from the simple fall-back picture.

Alternatively, GRB\,100316D could have signaled the birth of a magnetar.
With $E_{k}\sim 10^{52}\,\rm{erg}$ coupled to the non-relativistic ejecta 
(e.g. \citealt{Bufano12}), GRB\,100316D approaches
the limit of the magnetar models (Fig. \ref{Fig:Ekvelocity}).\footnote{For comparison,
as much as $\sim10^{54}\,\rm{erg}$ can be extracted from the black hole plus long-lived 
debris torus system (e.g. \citealt{Meszaros99}).}
For a maximally spinning, $1.4\,\rm{M_{\sun}}$ 
proto-neutron star the total energy release cannot exceed its rotation energy 
budget $E_{\rm{tot}}\sim 2.2\times 10^{52}\,\rm{erg}$
(e.g. \citealt{Thompson04}). In order to power a  $\sim 10^{52}\,\rm{erg}$ explosion, 
the highly-magnetized ($B\sim10^{15}\,\rm{G}$) compact object is required to be
born rapidly rotating (initial spin period $P\sim1\,\rm{ms}$), suggesting that 
GRB\,100316D is associated with the fastest and most extreme proto-magnetars.
While a significant fraction of rotation energy might be extracted in a few seconds
(e.g. \citealt{Metzger07}), the spin-down luminosity from the newly born magnetar 
is a source of long-lasting emission decaying as $L_p\propto t^{-l}$,
where $l=2$ corresponds to the magnetic dipole spin down. 
As in the case of GRB\,060218 (\citealt{Soderberg06b}), 
the shallow $L_x$ temporal decay argues for a spin-down braking index 
$n\approx 2$ (where $\dot\nu=-K \nu^n$ is the braking law
and $n=3$ is for the standard magnetic dipole radiation with constant magnetic field.
$\nu$ is the pulsar spin frequency).
Interestingly, a braking index $n<3$ has been observed in very young 
pulsars (e.g. \citealt{Livingstone07}).
The very soft X-ray spectrum could be either intrinsic or the result of the interaction 
of the central engine radiation with material ejected by the progenitor
(e.g. via multiple inelastic electron scatterings that suppress
high energy radiation). 




\section{Conclusions}
\label{Sec:conc}

Broad-band late-time monitoring of the sub-energetic GRB\,100316D
associated with SN\,2010bh  at radio and X-ray wavelengths  
allowed us to constrain the properties of the explosion and its
environment. The explosion is energetically dominated by 
the non-relativistic material which carries $E_k\sim 10^{52}\,\rm{erg}$. 
A  modest $E_k\sim 10^{49}\,\rm{erg}$ is coupled to quasi-spherical,  
mildly relativistic ($\Gamma= 1.5-2$) ejecta expanding into a medium
previously shaped by the progenitor mass-loss.
We infer a progenitor mass-loss rate of $\dot M \sim 10^{-5}\,\rm{M_{\sun}yr^{-1}}$, for an assumed 
wind velocity $v_{w}=1000\,\rm{km\,s^{-1}}$ and microphysical parameters
$\epsilon_e=0.01-0.1$ and $\epsilon_B=0.01$.

These properties are intermediate between the highly relativistic,
collimated GRBs and the spherical, ordinary hydrogen-stripped SNe.
Like GRBs, but different than ordinary Type Ib/c SNe, the 
kinetic energy profile of GRB\,100316D argues for the 
presence of a central engine that drives the explosion. However,
unlike GRBs, our broad-band spectral modeling
clearly identifies the presence of an excess of soft X-ray radiation
with respect to the synchrotron afterglow model at late times. This result leads us to conclude 
that the late-time ($t>10$ days) non-thermal radio and X-ray emission 
does not originate from the same population of electrons and are likely attributable
to two different components. 

We connect the excess of soft X-ray radiation to long-lasting activity of the explosion 
central engine, either in the form of a black hole plus torus system, or a magnetar.
The very high kinetic energy ($E_k\sim 10^{52}\,\rm{erg}$) carried by the non-relativistic
ejecta of GRB\,100316D implies that only the most rapidly rotating magnetars with spin period 
$P\sim1\,\rm{ms}$ can power the explosion. Accretion onto a compact object cannot
be excluded, but requires some significant departure from the standard
fall-back scenario.

GRB\,060218 is the only other explosion to date where a similar super-soft  X-ray 
excess has been identified at late times (\citealt{Soderberg06b, Fan06b,Waxman07}).
However, given the intrinsic faintness of the central engine component, 
it is possible that a similar emission is ubiquitous in long GRBs, but 
easily over-shone by the external shock afterglow associated with the highly relativistic jet.

Finally, GRB\,100316D and its cosmic twin, GRB\,060218, define a class of sub-energetic
explosions that are clearly distinguished from ordinary GRBs and other sub-energetic
GRBs (e.g. 120422A, Zauderer et al., in prep.) 
by: (i) a very long and smooth $\gamma$-ray prompt emission phase ($\Delta t>1000$ s)
with spectral peak energy $E_{\rm{pk}}\sim 10$ keV, (ii) shallower than average late-time 
decay of the X-ray light-curve, (iii) exceptionally soft late-time X-ray spectrum, and (iv)
evidence for an excess of soft X-ray emission with respect to the external shock afterglow
at late times. These two GRBs have also been claimed to show evidence for a thermal component in their 
early time X-ray afterglow (\citealt{Campana06,Starling11}). Our re-analysis however
points to a substantially reduced statistical significance of the black-body component in 
the early time spectra of GRB\,100316D.

Further progress in our understanding of central-engine driven explosions strongly
relies on the ability to constrain their energetics, collimation and environment properties. 
This is a task that can only be accomplished with coordinated efforts at radio and X-ray 
wavelengths at late times, when the central engine reveals itself in the X-rays.


\acknowledgments 
R.~M. is indebted to Cristiano Guidorzi for many interesting discussions.
R.~M. and B.~J.~M.  thank Dominic Ryan for useful conversations.
R.~M. thanks the KITP in Santa Barbara for their support, hospitality and the
stimulating environment that partially inspired this work. 
R.~B.~D. was supported by an ERC advanced grant (GRB) and by the I-CORE Program of the PBC and the ISF (grant 1829/12).
L.~S. is supported by NASA through Einstein Postdoctoral Fellowship 
grant number PF1-120090 awarded by the Chandra X-ray Center, which 
is operated by the Smithsonian Astrophysical Observatory for NASA under contract NAS8-03060.
B.~J.~M.  is supported by an NSF Astronomy and Astrophysics Post-doctoral Fellowship under award AST1102796. 
E.~P. acknowledges support from INAF PRIN 2011.
R.~A.~C. acknowledges support from NASA grant NNX12AF90G.
Support for this work was provided by the David and Lucile Packard Foundation Fellowship for Science and Engineering awarded
to A.~M.~S. 
The Australia Telescope Compact Array is part of the Australia Telescope National Facility which 
is funded by the Commonwealth of Australia for operation as a National Facility managed by CSIRO.


\end{document}